\documentclass{ifacconf}

\usepackage{graphicx}      
\usepackage{natbib}        
\usepackage{amsmath,amssymb}
\usepackage[algo2e,linesnumbered,ruled]{algorithm2e}
\usepackage{parskip}
\usepackage{subcaption}
\usepackage{epstopdf}
\usepackage{mathtools}

\newtheorem{theorem}{Theorem}

\newcommand{\vect}[1]{\ensuremath{\boldsymbol{\mathrm{#1}}}}
\newcommand{\secref}[1]{Section \ref{#1}}
\begin{document}
\begin{frontmatter}

\title{Learning-based MPC from Big Data Using Reinforcement Learning} 

\thanks[footnoteinfo]{This research was funded by the Research Council of Norway (RCN) through the project \textit{Safe Reinforcement Learning using MPC} (SARLEM), grant number 300172.}

\author[First]{Shambhuraj Sawant} 
\author[First]{Akhil S Anand} 
\author[First]{Dirk Reinhardt}
\author[First]{Sebastien Gros} 

\address[First]{Department of Engineering Cybernetics, Norwegian University of Science and Technology (NTNU) Trondheim, Norway \\
(e-mail: shambhuraj.sawant@ntnu.no, akhil.s.anand@ntnu.no, dirk.p.reinhardt@ntnu.no, sebastien.gros@ntnu.no).}

\begin{abstract}
This paper presents an approach for learning Model Predictive Control (MPC) schemes directly from data using Reinforcement Learning (RL) methods. The state-of-the-art learning methods use RL to improve the performance of parameterized MPC schemes. However, these learning algorithms are often gradient-based methods that require frequent evaluations of computationally expensive MPC schemes, thereby restricting their use on big datasets. We propose to tackle this issue by using tools from RL to learn a parameterized MPC scheme directly from data in an offline fashion. Our approach derives an MPC scheme without having to solve it over the collected dataset, thereby eliminating the computational complexity of existing techniques for big data. We evaluate the proposed method on three simulated experiments of varying complexity.

\end{abstract}

\begin{keyword}
Big Data, Model Predictive Control, Reinforcement Learning
\end{keyword}

\end{frontmatter}


\section{Introduction}
Model Predictive Control (MPC) is a pervasive control methodology in modern industrial, power, and robotics applications. It is based on a well-established theory to handle multivariate dynamics by generating sequences of optimal control inputs that minimize a cost function, possibly subject to constraints \citep{rawlings2017mpc}. The common practice for designing MPC schemes includes identifying a model of the system dynamics using the collected data and specifying the shape of a cost function that encapsulates the control objective. Finding a model that accurately fits the true dynamics often requires much work, particularly for systems with complex dynamics. In many cases, it might be necessary to reduce model complexity such that an embedded computing platform can meet the computational demands and memory footprint of the resulting MPC scheme. Furthermore, even after finding a suitable model, tuning the cost function can be cumbersome for tracking problems and even harder for economic tasks. In summary, choosing an optimal set of parameters for a given MPC design usually demands an expert's time and effort. 

Many recent approaches for improving the performance of MPC schemes focus on data-driven machine learning techniques for adjusting the model to better fit the system dynamics \citep{wu2019machine, lucia2018deep}. The reasoning is that the predictive qualities of the underlying model are the main bottleneck for achieving optimal performance. However, this approach may not necessarily yield better closed-loop performance as the model adjustment is not tied directly to performance improvement. 
Furthermore, it is possible to obtain further performance gains by additionally adjusting the associated cost and constraints. 

\citep{gros2019data} formally justified this idea in the context of Reinforcement Learning (RL) for MPC by introducing MPC as a function approximator for the RL. Their MPC-based RL approach learns MPC parameters directly using RL to improve the performance rather than to improve the prediction accuracy of the model. In a fully parameterized MPC formulation, RL can be used to learn the cost, model, and constraint functions, thereby providing high flexibility in function approximation. 
This approach can also be thought of as tuning MPC parameters using RL tools for improved performance. 
In recent times, various works were presented investigating the MPC-based RL approach and their use in various applications \citep{cai2021mpc, kordabad2021mpc, cai2021optimal}.  

The existing MPC-based RL methods iteratively improve the MPC parameters in an online fashion, by using the system behaviour under the current MPC policy. These methods involve evaluating an MPC scheme and the associated sensitivities to compute the gradient step required for the parameter update. These required MPC and sensitivity evaluations can be computationally demanding. This problem gets further exasperated when carrying out batch parameter updates. Additionally, the existing methods require access to the system for on-policy interactions. Consequently, using the existing MPC-based RL methods over large datasets gets challenging. We provide further discussion on the involved challenges in \secref{sec:bigdata_challenges}. 
Learning an MPC scheme based on collected data at a low computational expense before interacting with the system should appeal to any MPC practitioner and we take a step toward enabling this in the current work. 
This task of learning a policy from previously collected data is also referred to as \textit{offline} RL in the RL literature \citep{levine2020offline} and several such methods have been proposed to effectively learn from offline datasets \citep{kumar2020conservative, kidambi2020morel, wu2019behavior, agarwal2020optimistic}.

In this paper, we propose a novel approach for learning a performance-oriented MPC formulation in an offline RL setting from the previously collected data. We leverage our approach on \citep[Theorem 1]{gros2019data}, which shows the relation between the optimal MPC parameters and the optimal value function of the underlying system.
Our approach makes use of this connection to learn a high-performing MPC scheme from data while avoiding expensive MPC evaluations. With our approach, the task of learning a high-performing MPC scheme from data is reduced to a supervised learning problem and it is solved at a low computational expense as compared with the existing MPC-based RL methods. An added benefit of our approach is that it learns an MPC policy without having to evaluate any MPC scheme.

The rest of the paper is organized as follows: Section \ref{sec:back} briefly introduces the necessary background knowledge and the challenges present in using the existing MPC-based RL methods on big data, while Section \ref{sec:method} details our proposed method for learning an MPC scheme from data. Section \ref{sec:exp} contains the evaluation of our approach on three different simulation experiments of varying complexity. A brief discussion of the evaluation results of our method and further challenges is presented in Section \ref{sec:discussion} and conclusions are discussed in Section \ref{sec:con}.
\section{Background}\label{sec:back}

\subsection{Markov Decision Process}
Markov Decision Processes (MDPs) provide a generic framework for the class of problems at the core of MPC. An MDP operates over given state and action (aka input) spaces $\mathcal{S},\,\mathcal{A}$, respectively. These spaces can be discrete (i.e. integer sets), continuous, or mixed. An MDP is defined by the tuple $(\mathcal{S},\, \mathcal{A},\,L,\,\gamma,\,\rho)$, where $L$ is a stage cost, $\gamma\in[0,1)$ a discount factor, and 
$\rho$ denotes the conditional probability (measure) defining the dynamics of the underlying system, i.e. for a given state-action pair $\vect s,\,\vect a\in \mathcal{S}\times \mathcal{A}$, the successive state $\vect s_+$ is distributed according to 
\begin{align}
\label{eq:Dyn} 
\vect s_+ \sim \rho(\cdot |\vect s,\vect a) \, .
\end{align}
Note that \eqref{eq:Dyn} is a generalization of the deterministic or stochastic dynamics model often considered in MPC, usually cast as:
\begin{align}
\label{eq:Dyn:Classic} 
\vect s_+ = \vect F\left(\vect s,\vect a,\vect w \right),\quad \vect w\sim W
\end{align}
where $\vect w$ is a random disturbance from distribution $W$. In the special case $\vect w=0$, \eqref{eq:Dyn:Classic} simply yields deterministic dynamics. A solution to an MDP then involves finding an optimal policy $\vect \pi^\star\,:\, \mathcal{S}\rightarrow \mathcal{A}$, given as:
\begin{equation}
\vect \pi^\star = \mathrm{arg}\min_{\vect\pi}\, J(\vect \pi) \label{eq:OptPolicy:Def}
\end{equation}
where $J(\vect \pi)$ is a measure of the closed-loop performance which is defined as the cumulative cost, i.e.:
\begin{equation}
J(\vect \pi) = \mathbb{E}\left[\left. \sum_{k=0}^\infty\, \gamma^k L\left(\vect s_k,\vect a_k\right)\,\right|\, \vect a_k=\vect \pi\left(\vect s_k\right)\right] \, .
\end{equation}
The expected value operator $\mathbb{E}[.]$ is taken over the (possibly) stochastic closed loop trajectories of the system. Discussing the solution of an MDP is often best done via the Bellman equations defining implicitly the optimal value function $V^\star\,:\, \mathcal{S}\rightarrow \mathbb R$ and the optimal action-value function $Q^\star\,:\, \mathcal{S}\times \mathcal{A}\rightarrow \mathbb R$ \citep{sutton2018reinforcement} as:
\begin{subequations}
\label{eq:Bellman0} 
\begin{align}
V^\star\left(\vect s\right) &=\min_{\vect a}\, Q^\star\left(\vect s,\vect a\right) \label{eq:Bellman0:V} \\
Q^\star\left(\vect s,\vect a\right) &= L\left(\vect s,\vect a\right) + \gamma \mathbb E\left[V^\star\left(\vect s_+\right)\,|\,\vect s,\vect a\, \right] 
\end{align}
\end{subequations}
The optimal policy then reads as:
\begin{align}
\label{eq:OptPolicy0} 
\vect \pi^\star\left(\vect s\right) =\mathrm{arg} \min_{\vect a}\, Q^\star\left(\vect s,\vect a\right) 
\end{align}

\subsection{Model Predictive Control}\label{sec:MPCIntro}
For a given system state $\vect s$, an MPC produces control policies based on repeatedly solving an optimal control problem on a finite, receding horizon. Suppose $\vect{f}_{\vect{\theta}}$ denotes a model of the true dynamics~\eqref{eq:Dyn}, $L_{\vect{\theta}}$ is the stage cost function, and $\vect h_{\vect{\theta}}$ denotes the constraints, each parameterized by a parameter vector $\vect \theta$. The problem to be solved is then typically cast as:
\begin{subequations}
\label{eq:MPC} 
\begin{align}
\min_{\vect x,\vect u}&\quad T_{\vect{\theta}} \left(\vect x_N\right) + \sum_{k=0}^{N-1}\, L_{\vect{\theta}}\left(\vect x_k,\vect u_k\right) \label{eq:MPC:Cost}\\
\mathrm{s.t.} &\quad \vect x_{k+1} = \vect f_{\vect{\theta}}\left(\vect x_k,\vect u_k\right),\quad \vect x_0 = \vect s \label{eq:MPC:Dyn}\\
&\quad \vect h_{\vect{\theta}}\left(\vect x_k,\vect u_k\right) \leq 0,\quad \vect u_k \in \mathcal{A}
\label{eq:MPC:Const}
\end{align}
\end{subequations}
For an initial condition $\vect x_0 = \vect s$, problem \eqref{eq:MPC} produces a sequence of control inputs $\vect u^\star = \{\vect u_0^\star,\ldots, \vect u_{N-1}^\star\}$ and corresponding state predictions $\vect x^\star= \{\vect x_0^\star,\ldots, \vect x_{N}^\star\}$. Only the first element $\vect u_0^\star$ of the input sequence $\vect u^\star$ is applied to the system. At the next sampling step, a new state $\vect s$ is received, and problem \eqref{eq:MPC} is solved again, producing a new $\vect u^\star$ and a new $\vect u_0^\star$. MPC hence yields a policy:
\begin{align}
\label{eq:MPC:Policy} 
\vect\pi_\theta\left(\vect s\right) = \vect u_0^\star\, ,
\end{align} 
with $\vect u_0^\star$ solution of \eqref{eq:MPC} for a given initial state $\vect s$. For $\gamma\approx 1$, policy \eqref{eq:MPC:Policy} can provide a good approximation of the optimal policy $\vect\pi^\star$ for an adequate choice of prediction horizon $N$, terminal cost $T_{\vect{\theta}}$ and if the MPC model $\vect{f}_{\vect{\theta}}$ approximates the true dynamics \eqref{eq:Dyn} sufficiently well. Still, the problem \eqref{eq:MPC} can be trivially extended for an underlying MPC with a discount factor $\gamma$ as:
\begin{subequations}
\label{eq:MPCd} 
\begin{align}
\min_{\vect x,\vect u}&\quad \gamma^N T_{\vect{\theta}} \left(\vect x_N\right) + \sum_{k=0}^{N-1}\, \gamma^k L_{\vect{\theta}}\left(\vect x_k,\vect u_k\right) \label{eq:MPCd:Cost}\\
\mathrm{s.t.} &\quad \eqref{eq:MPC:Dyn} - \eqref{eq:MPC:Const} \label{eq:MPCd:Const}
\end{align}
\end{subequations}

In the context approximating $\vect \pi^\star(\vect s)$ using \eqref{eq:MPC:Policy}, the model is arguably the major bottleneck as many systems are challenging to model accurately. Machine learning approaches for MPC that modify the model based on mismatch to the observations, e.g. using Gaussian Process regression, are precisely tackling this problem. However, within a modeling structure, selecting the model $\vect f_{\vect{\theta}}$ that yields the best closed-loop performance $J(\vect\pi_{\vect{\theta}})$ is very difficult. Additionally, there is in general no guarantee that the most accurate model yields the best closed-loop performance. 

\subsection{Reinforcement Learning for MPC}

It is essential to understand how the optimal value functions and policy can be approximated by an MPC scheme. Using \eqref{eq:MPCd}, we consider an approximation of the value function as
\begin{subequations}
\label{eq:MPC:V}
\begin{align}
V_{\vect{\theta}}(\vect s) = \min_{\vect x,\vect u}&\quad \eqref{eq:MPCd:Cost},\\
\mathrm{s.t.} &\quad \eqref{eq:MPC:Dyn} - \eqref{eq:MPC:Const} \label{eq:MPC:Q:II}\,
\end{align}
\end{subequations}
and the action-value function as
\begin{subequations}
\label{eq:MPC:Q}
\begin{align}
Q_{\vect{\theta}}(\vect s,\vect a) = \min_{\vect x,\vect u}&\quad \eqref{eq:MPCd:Cost},\\
\mathrm{s.t.} &\quad \eqref{eq:MPC:Dyn} - \eqref{eq:MPC:Const},\quad \vect u_0 = \vect a \label{eq:MPC:Q:II}\,
\end{align}
\end{subequations}

with added constraint $\vect u_0 = \vect a$ on the initial input. The MPC scheme in \eqref{eq:MPC:Q} is a valid approximation of $Q^\star$ in the sense that it satisfies the relationships \eqref{eq:Bellman0} and \eqref{eq:OptPolicy0}, i.e.:
\begin{align}
\label{eq:MPC:PiV}
\vect\pi_{\vect{\theta}}\left(\vect s\right) &= \mathrm{arg}\min_{\vect a} Q_{\vect{\theta}}(\vect s,\vect a),\quad
V_{\vect{\theta}}(\vect s) =\min_{\vect a}\,Q_{\vect{\theta}}(\vect s,\vect a)\,.
\end{align}

For learning the approximations in \eqref{eq:MPC:V} and \eqref{eq:MPC:Q}, let us briefly state the central result by \cite{gros2019data}. It establishes the equivalence between the optimal policy and value functions and the approximations provided by the MPC:

\begin{theorem}[{{\cite{gros2019data}}}] \label{th:rlmpc}
Consider the parameterized stage cost, terminal cost and constraints in \eqref{eq:MPCd} as function approximators with adjustable parameters $\vect\theta$. Further suppose that $\vect x^\star$ is an optimal state trajectory generated by the MPC in \eqref{eq:MPCd} and there exist parameters $\vect\theta^\star$ such that 
\begin{subequations}\label{eq:st2}%
\begin{align}%
T_{\vect\theta^\star}(\vect s) &= V^\star (\vect s) \label{eq:That0}\\
  L_{\vect\theta^\star}(\vect s,\vect a) &= \left\{\begin{matrix}
 Q^\star(\vect s,\vect a)-\gamma V^+(\vect s, \vect a) & \mathrm{If}\, \left|V^+(\vect s, \vect a)\right|<\infty \\ \infty
 & \mathrm{otherwise}
\end{matrix}\right.\label{eq:lhat0}
\end{align}
\end{subequations}
where, $V^+(\vect s, \vect a) = V^\star(\vect f_{\vect\theta^\star}(\vect s,\vect a))$. Then the following identities hold $\forall\, \gamma$:
\begin{enumerate}
  \item $V_{\vect\theta^\star}(\vect s)=V^\star(\vect s),\,\forall \vect s\in\mathcal{S}$\label{eq:VV}
  \item ${\vect\pi}_{\vect\theta^\star}(\vect s)=\vect\pi^\star(\vect s),\,\forall \vect s\in\mathcal{S}$
  \item $ Q_{\vect\theta^\star}(\vect s,\vect a)=Q^\star(\vect s,\vect a),\,\forall \vect s\in\mathcal{S}$, for the inputs $\vect a\in \mathcal{A}$ such that $\lvert V^\star(\vect f_{\vect\theta^\star}(\vect s,\vect a))\rvert<\infty$\label{eq:QQ}
\end{enumerate}
if the set 
\begin{align}\label{eq:assum:Theo}
\Omega&=:\left\{\vect s\in \mathcal{S}\,\,\Big|\,\,\left| \left[V^\star(\vect x^\star_k)\right]\right|<\infty, \ \forall\, k \leq  N\right\}
\end{align}
is non-empty.
\end{theorem}
The proof follows from \citep{gros2019data, kordabad2022equivalency}.



Essentially, Theorem \ref{th:rlmpc} states that it is possible to compute the optimal policy $\vect \pi^\star(\vect s)$ using an inaccurate model dynamics. However, $L_{\theta^\star}(\vect s, \vect a)$ in \eqref{eq:lhat0} is difficult to compute and requires a model of the system dynamics. To bypass difficult $L_{\theta^\star}(\vect s, \vect a)$ evaluation, \citep{gros2019data} suggest to learn the optimal parameter $\theta^\star$ 
using RL tools such as $Q$ learning or policy gradient methods.

\subsection{Challenges with learning MPC with RL on big data}\label{sec:bigdata_challenges}
Most RL methods involve iteratively optimizing the policy using the experience acquired by interacting with the system. Their success hinges on repeatedly combing through collected data to build better approximations of MDP value functions and policy, in particular when using Deep Neural Networks (DNNs) as function approximators. The MPC-based RL methods, similarly, iteratively optimize the closed-loop performance for learning $\vect{\theta}^*$ from observed state transitions. However, these iterative updates require computing multiple MPC solutions for each optimization step. Additionally, finding an optimal $\vect{\theta}^*$ by scrubbing through the collected data is computationally demanding.

The classical RL methods fundamentally operate in \textit{online} learning paradigm. Though many RL methods work with off-policy data, these methods often cannot learn effectively from entirely \textit{offline} datasets, without additional on-policy interactions \citep{levine2020offline}. In recent times, the offline RL paradigm has garnered growing interest and many approaches have been proposed to effectively learn from offline data \citep{kumar2020conservative, kidambi2020morel, wu2019behavior, agarwal2020optimistic}. However, many challenges persist in using RL with offline data as discussed in \citep{levine2020offline, fu2020d4rl}.
\section{Learning MPC from big data}\label{sec:method}
In this section, we describe our approach to learning an MPC scheme from big data. 

\subsection*{Learning MPC parameterizations}
As stated in Theorem \ref{th:rlmpc}, under some conditions, an MPC scheme using a model $\vect f_{\vect{\theta}}(\vect s,\vect a)$ and the stage cost in \eqref{eq:lhat0} yields the optimal policy $\vect \pi^\star(\vect s)$ for the underlying MDP, together with the associated optimal value functions \eqref{eq:Bellman0}. The optimal parameters for an MPC scheme parameterized in $\vect \theta$ relates to a well-defined optimal value function $V^\star (\vect s)$, as in given in Theorem \ref{th:rlmpc}, as:
\begin{align}
    L_{\vect \theta^*}(\vect s,\vect a) =&\ Q^\star(\vect s,\vect a)-\gamma V^\star(\vect f_{\vect \theta^*}(\vect s,\vect a)) \, .
\end{align}
Using Bellman equations, we get, 
\begin{align}\label{eq:st3}
    L_{\vect \theta^*} (\vect s, \vect a) =&\ L(\vect s, \vect a) + \gamma V^\star(\vect s_+)-\gamma V^\star(\vect f_{\vect \theta^*}(\vect s,\vect a)) \, ,
\end{align}
where $L(\vect s, \vect a)$ is the stage cost of the underlying MDP. Here, variable $\vect\theta$ includes all the parameterizations introduced in the MPC scheme in \eqref{eq:MPC}.

Estimating the optimal value function $V^\star(\vect s)$ requires knowledge of the system dynamics \eqref{eq:Dyn} and the stage cost $L(\vect s, \vect a)$. However, $V^\star(\vect s)$ can also be approximated from a dataset with a rich enough data distribution over the state and action space $\mathcal{S}, \mathcal{A}$. We propose to use Deep Neural Networks (DNNs) to approximate a value function $V_{\vect \phi}(\vect s) \approx V^*(s)$ from data. $V_{\vect \phi}(\vect s)$ can be estimated using update equation given as: 
\begin{align}
    \vect \phi^* &= \arg\min_{\vect \phi} \mathbb{E}_{\tau \sim \mathcal{D}} \left[ (L(\vect s, \vect a) + \gamma V_{\vect \phi}(\vect s_+) - V_{\vect \phi}(\vect s) )^2 \right] \, ,
\end{align}
where $\tau: \{\vect s,\vect a,\vect s_+,L(\vect s,\vect a)\}$ is the transition sampled from the given dataset $\mathcal{D}$. 
With a dataset $\mathcal{D}$ generated following a greedy in the limit of infinite exploration (GLIE) policy, the learned value function $V_{\vect \phi}(\vect s)$ will closely match match the optimal value function $V^\star(\vect s)$ i.e. $V_{\vect \phi}(\vect s) \sim V^\star(\vect s)$. 
The estimated $V_{\vect\phi}(\vect s)$ can then be integrated into \eqref{eq:st3} as: 
 \begin{align}\label{eq:st4}
    {L}_{\vect \theta^*} (\vect s, \vect a) \approx &\ L(\vect s, \vect a) + \gamma V_{\vect \phi}(\vect s_+)-\gamma V_{\vect \phi}(\vect f_{\vect \theta}(\vect s,\vect a)) \, .
\end{align}
 

It is often useful to impose additional constraints on the parameterization adopted in an MPC scheme, e.g. the stage cost should preferably be convex and smooth, such that $L_{\vect \theta}$ may preferably be restricted to a specific class of functions. Considering possible constraints on the stage cost, we fit a structured cost $\hat{L}_{\vect \theta}(\vect s, \vect a)$ to the cost function $L_{\vect \theta^*} (\vect s, \vect a)$ in \eqref{eq:st4}. Thus, the optimal parameters $\vect \theta^*$ for the parameterized MPC scheme can be derived as:
\begin{subequations}
\begin{align}\label{eq:update_rule}
    \quad \vect \theta^* = \arg\min_{\vect \theta} \mathbb{E} &[ (L(\vect s, \vect a) + \gamma V_{\vect\phi}(\vect s_+) \nonumber \\
    & - \gamma V_{\vect\phi}(\vect f_{\vect \theta}(\vect s, \vect a))- \hat{L}_{\vect \theta} (\vect s,\vect a))^2 ] 
\end{align} 
\end{subequations}
This forms the central result of our work, which essentially reduces the problem of learning MPC parameters from data to a supervised learning problem. 

An important observation regarding $\vect \theta^*$ from \eqref{eq:st2} and \eqref{eq:update_rule} is that the resulting model dynamics $\vect f_{\vect \theta^*}(\vect s, \vect a)$ may not match the true system dynamics. The model dynamics $\vect f_{\vect \theta^*}(\vect s, \vect a)$ is unlikely to be the best fit model for state predictions as it is solely adjusted for closed-loop performance. However, if it is important to have a model with a high prediction accuracy, the parameters associated with the model dynamics $\vect f_{\vect \theta}(\vect s, \vect a)$ can be excluded from $\vect \theta$ i.e. minimizing \eqref{eq:update_rule} over only cost parameterizations. Indeed, according to Theorem \ref{th:rlmpc}, if one can accommodate a rich structure to the stage cost, then the MPC formulated by learning only the stage cost using \eqref{eq:update_rule} can compensate for any model inaccuracies. We will discuss further the issue of the limitations on the structure of the cost function $\hat{L}_{\vect \theta}$ and some possible solutions in Section \ref{sec:discussion}. Additionally, Theorem \ref{th:rlmpc} holds for optimal value function $V^\star(\vect s)$ which can be built given a dataset generated with a greedy in the limit of exploration (GLIE) policy \citep{sutton2018reinforcement}. However, that is a fair concern for all data-driven learning methods. In our approach, with a rich dataset, a good approximation of the value function can still be built and used for learning a high-performing MPC scheme. 



To summarize, we propose a solution to formulate a high-performing MPC scheme from big data using RL tools. We reduce the problem of learning the MPC parameters to a supervised learning task by building a value function estimate from a given dataset and incorporating it into Theorem \ref{th:rlmpc}. To that extent, we utilize the strength of machine learning methods to extract knowledge from big data effectively and utilize it for learning the MPC scheme, thereby eliminating the enormous computational complexities faced by existing approaches for learning-based MPC. As we outsource the problem of extracting a value function estimate and learning MPC parameters to machine learning tools, the learned MPC scheme is obtained at low computational expense without any MPC evaluations. Additionally, it is assured to have a high closed-loop performance if a good approximation of $V^\star(\vect s)$ can be built from the data and the learned MPC parameters satisfy \eqref{eq:update_rule}. In the next section, we will evaluate the proposed approach in three different simulation experiments.

\section{Experiments}\label{sec:exp}

In this section, we present three simulated examples of varying complexity to evaluate our approach to learning MPC from big data. 

\subsection{Experimental Setup}
\subsubsection{MPC parameterization:} For all the three simulated examples, we parameterized modified stage cost $\hat{L}_{\vect \theta}(\vect s, \vect a)$ in \eqref{eq:update_rule} as quadratic cost given by
\begin{align} 
    \hat{L}_{\vect \theta}(\vect s,\vect a) &= (\vect s-\vect s_{ref})^T \vect W_{\vect \theta} (\vect s-\vect s_{ref}) + \vect a^T \vect R_{\vect \theta} \vect a + \theta_3 \label{eq:st5} \\
    \vect W_{\vect \theta} &= \begin{bmatrix}
\theta_{W,11} & \theta_{W,12} & \dots \\
\theta_{W,21} & \theta_{W,22} & \dots \\
\vdots & \vdots & \ddots \\
\end{bmatrix},\,
    \vect R_{\vect \theta} = \begin{bmatrix}
\theta_{R,11} & \theta_{R,12} & \dots \\
\theta_{R,21} & \theta_{R,22} & \dots \\
\vdots & \vdots & \ddots \\
\end{bmatrix} \nonumber \, ,
\end{align}
where $\vect W_{\vect \theta}, \vect R_{\vect \theta}$ are the state and input weight matrices, and $\vect s_{ref}$ is the desired goal state of the tracking control problem. The terminal cost is set to be $2$ norm of tracking error, $T_{\vect \theta}(s) = \lVert \vect s - \vect s_{ref} \rVert$. The parameter vector $\vect \theta$ collects all the parameters in the MPC scheme together. In order to formulate a well-defined MPC scheme, we further include a semi-definite constraint over $\vect W_{\vect \theta}$ and $\vect R_{\vect \theta}$. Additionally, $l2$ regularization penalty is imposed around the given initial parameter estimates. The $l2$ regularization is used to restrict the RL updates on the parameters to be closer to their initial values, thereby facilitating stable updates. We choose to fix the constraint equations for all the experiments i.e. constraints are not parameterized. The three different tracking control problems chosen to test our method are
\begin{enumerate}
    \item Linear tracking MPC task,
    \item Pendulum swing-up task,  
    \item Cartpole swing-up and balancing task.
\end{enumerate}

In all the experiments, the variable $\vect \theta$ is initialized to a nominal MPC scheme formulated using an inaccurate model and a nominal stage cost. Essentially we apply our approach to improve the performance of this nominal MPC formulation directly from a dataset. The nominal MPC parameters can be seen as an initial guess for our approach to improve upon. This nominal MPC is denoted by $\mathit{MPC}_1$ throughout this section. 

In all three experiments, we learn two different MPC parameterizations,
\begin{enumerate}
    \item Only learning the stage cost parameters with a fixed model: denoted as  $\mathit{MPC}_2$ where the $\vect \theta$ in \eqref{eq:update_rule} only constitute the stage cost parameters in \eqref{eq:st5}. 
    \item Learning both stage cost and model parameters: denoted by $\mathit{MPC}_3$ where the $\vect \theta$ constitutes both the stage cost parameters in \eqref{eq:st5} and model parameters.
\end{enumerate}

The richer parameterization in the second case allows higher flexibility for capturing $L_{\vect \theta}(\vect s, \vect a)$ in \eqref{eq:st3}. However, learning cost parameterization with fixed model dynamics should still capture \eqref{eq:st3} to a degree and learn a better performing MPC scheme than the nominal one. In order to evaluate our approach we compare the performance achieved by both of these learning-based MPC schemes to $\mathit{MPC}_1$ in the case of pendulum swing-up and cartpole swing-up tasks. For the linear tracking MPC task, the performance for these MPC schemes is reported relative to closed-loop optimal performance. 

\subsubsection{Data Generation:} 
A rich dataset was obtained using a popular RL method, Deep-Deterministic Policy Gradient ($\mathit{DDPG}$) \citep{lillicrap2015continuous}. For each example, a dataset of $500$ episodes (each with $100$ transitions) is collected using $\mathit{DDPG}$ agent. The $\mathit{DDPG}$ agent's learning progress ensures rich data distribution in the collected dataset.
\subsubsection{Value Learning:}
For all three experiments, we learn the approximations for value functions $V_{\vect \phi}(\vect s)$ using DNNs with 2 hidden layers (each with 256 neurons).  With current machine learning tools, building a good approximation of $V^\star(\vect s)$ from big data takes a fraction of computational effort compared to that of learning based MPC methods. 

\subsection{Linear Tracking MPC}
We first consider a simple linear MPC to illustrate the effectiveness of our approach to learning an MPC scheme from data. The linear MPC task consists of a four-dimensional state space, $\vect s = [x,y,\dot{x},\dot{y}]$ and input space, $\vect a = [F_x, F_y]$. The true system dynamics is of deterministic nature, defined as:
\begin{align}
\vect s_+ =&\ \vect A\, \vect s+ \vect B\, \vect a \label{eq:pointmass}\\
\vect A =&\ \begin{bmatrix}
1 & 0 & 0.1 & 0 \\
0 & 1 & 0 & 0.1 \\
0 & 0 & 0.9 & 0 \\
0 & 0 & 0 & 0.9
\end{bmatrix} \nonumber \\
&\ + \alpha \begin{bmatrix}
0.68 & -1.15 & -2.29 & -2.42 \\
1.57 &  2.06 &  0.53 &  1.15 \\
0.22 &  2.17 &  1.58 & -2.49 \\
1.79 & -2.33 &  1.15 & -1.62
\end{bmatrix}\times10^{-2} \nonumber\\
\vect B =&\ \begin{bmatrix}
0 & 0 \\
0 & 0 \\
0.1 & 0 \\
0 & 0.1
\end{bmatrix} +\alpha \begin{bmatrix}
1.82 &  0.21 \\
-1.00 & -0.39 \\
-2.36 & -1.88 \\
0.85 &  0.74
\end{bmatrix}\times10^{-2} \nonumber
\end{align}
where $\alpha$ scales the unmodeled part of system dynamics. 
Note that, by varying the value of $\alpha$ we can generate different system dynamics. The stage cost for the true system is given as 
\begin{align*}
    L(\vect s, \vect a) &= 9 (x^2+y^2)+1(\dot{x}^2+\dot{y}^2)+0.1(F_x^2+F_y^2) \,.
\end{align*}
The task is discretized with sampling time of $0.1$, the discount factor $\gamma$ is $0.9$, and the task length is $100$. The MPC scheme is defined for discretized control task with a horizon equal to the task length i.e. $N=100$. $\mathit{MPC}_1$ is formed using $L(\vect s, \vect a)$ as the stage cost and nominal model dynamics with $\alpha = 0$ in \eqref{eq:pointmass}. 
For $\mathit{MPC}_2$, the stage cost is parameterized by $\hat{L}_{\vect \theta}(\vect s, \vect a)$ in \eqref{eq:st5} and a nominal model dynamics with $\alpha=0$ in \eqref{eq:pointmass}. For $\mathit{MPC}_3$, the MPC scheme is formed using $\hat{L}_{\vect \theta}(\vect s, \vect a)$ in \eqref{eq:st5} as a stage cost and a model dynamics with $(\vect A_{\vect \theta}, \vect B_{\vect \theta})$, where $(\vect A_{\vect \theta}, \vect B_{\vect \theta})$ are fully parameterized matrices of corresponding sizes.

\begin{figure}[h]
\centering
\includegraphics[width=7cm]{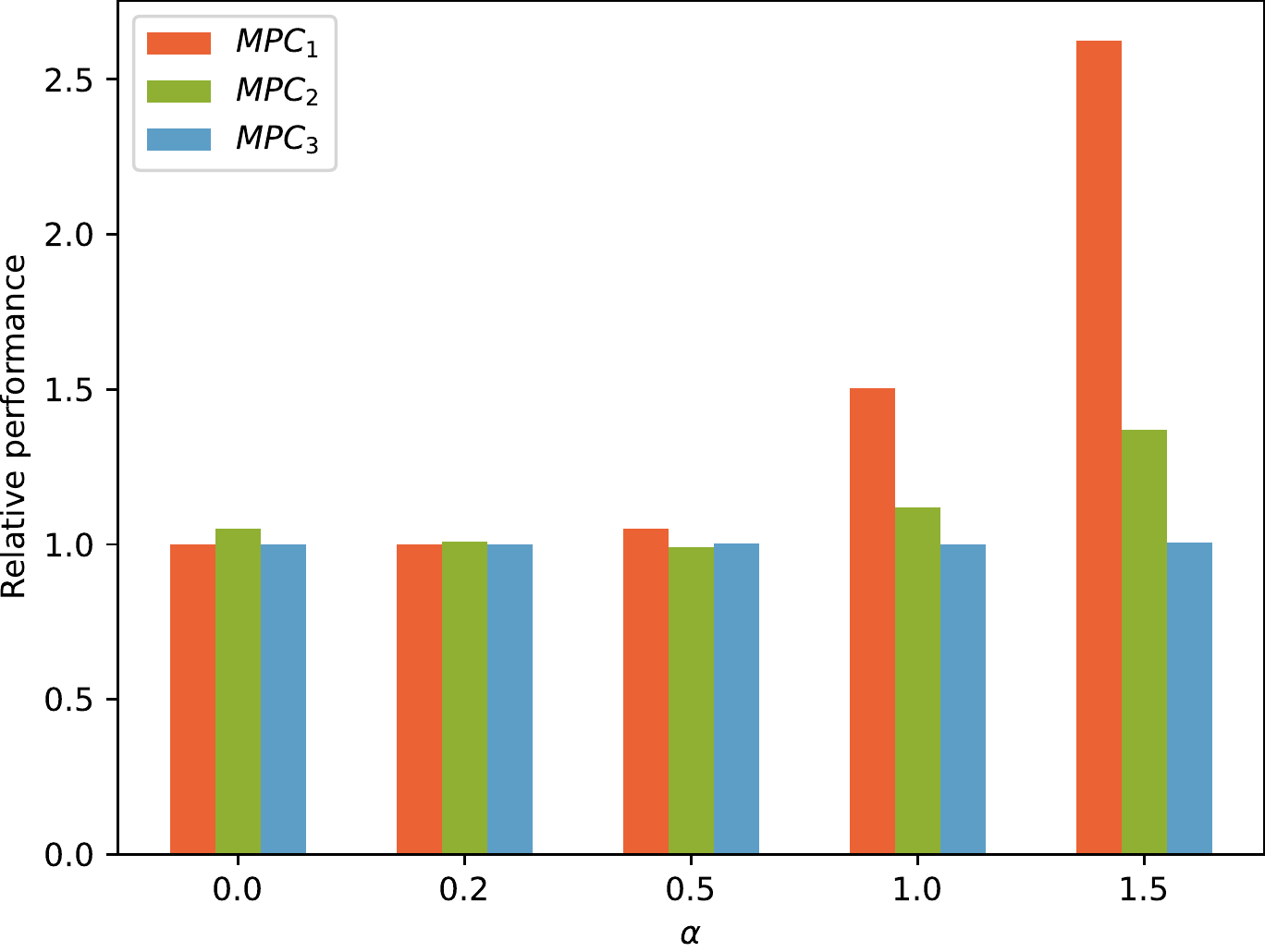}
\caption{Performance of the three MPC schemes relative to the closed-loop optimal performance for a linear tracking MPC task.}
\label{fig:perf_pointmass}
\end{figure}

Figure \ref{fig:perf_pointmass} shows the relative performance of MPC schemes with respect to closed-loop optimal performance for the system dynamics corresponding with different values of $\alpha$. 
As seen from fig. \ref{fig:perf_pointmass}, performance of $\mathit{MPC}_1$ starts to degrade with added perturbations, as the plant-model mismatch increases. Learning cost parameterization in $\mathit{MPC}_2$ manages to find a high-performing MPC scheme for the inaccurate model dynamics, i.e. the learnt cost parameters compensate for the model inaccuracies for lower value of $\alpha$. However, as $\alpha$ increases, $\mathit{MPC}_2$ with a quadratic cost parameterization in \eqref{eq:st5} fails to capture $L_{\vect \theta^*}(\vect s, \vect a)$ in \eqref{eq:st3}, resulting in loss of performance. Whereas, $\mathit{MPC}_3$ with learned cost and model parameterizations finds a high performing MPC scheme for all value of $\alpha$ i.e. with our approach, an MPC scheme with a close-to-optimal performance can be learned from the collected data.

\subsection{Pendulum Swing-Up}
We consider a pendulum swing-up task to illustrate our method on non-linear dynamics with a complex value function. In this example, the goal of the control task is to swing up the pendulum with an under-actuated motor and maintain it at an upright angle. The system dynamics are given by
\begin{align}
    \ddot{\phi} = \frac{3g}{2l}\sin{\phi}+\frac{3}{ml^2}u
\end{align}
where $\phi$ is the angular deviation of the pole w.r.t the vertical position, gravity $g=9.8$, $u$ is the applied torque, $u \in [-2,2]$, and $m=1,\, l=1$ are the mass and length of the pole respectively. The system state $\vect s$ is defined to be $\vect s=[\phi, \dot{\phi}]$ with $\phi$ normalized between $(-\pi,\pi]$. The corresponding cost function is given as
\begin{align*}
    L(\vect s, u) = \phi^2+0.1\dot{\phi}^2+0.1u^2 \,.
\end{align*}
The task is discretized with a sampling time of $0.05$, the discount factor $\gamma$ is $0.9$, and the task length is $100$.

The MPC scheme for the discretized control task is defined over a transformed state-space $\vect y=[\cos{\phi},\sin{\phi},\dot{\phi}]$ solely for the ease of implementation. The horizon for the MPC scheme is chosen as $N=50$ and the tracking goal corresponding to the vertical pole orientation is $\vect y_{ref} = [1,0,0]$. The nominal stage cost for the MPC scheme over observation $\vect y$ is
\begin{align*}
    L_{\mathit{MPC}}(\vect y, u) = (\cos\phi -1)^2 + \sin \phi ^2 + 0.1\dot{\phi}^2 + 0.1 u^2 \,.
\end{align*}

We use an inaccurate model with an uncertain estimate of true system properties for forming the nominal MPC scheme $\mathit{MPC}_1$. The uncertain estimates of mass and length used in the MPC model are $\hat{m}$ and $\hat{l}$ respectively, with
\begin{align*}
    \hat{m} = m + \alpha\ \mathcal{U}[-0.5,0.5],\quad \hat{l} = l + \alpha\ \mathcal{U}[-0.5,0.5]
\end{align*}
where $\alpha$ scales the uncertainty in estimates of the mass and length of the pole.

\begin{figure}[h]
\centering
\includegraphics[width=7cm]{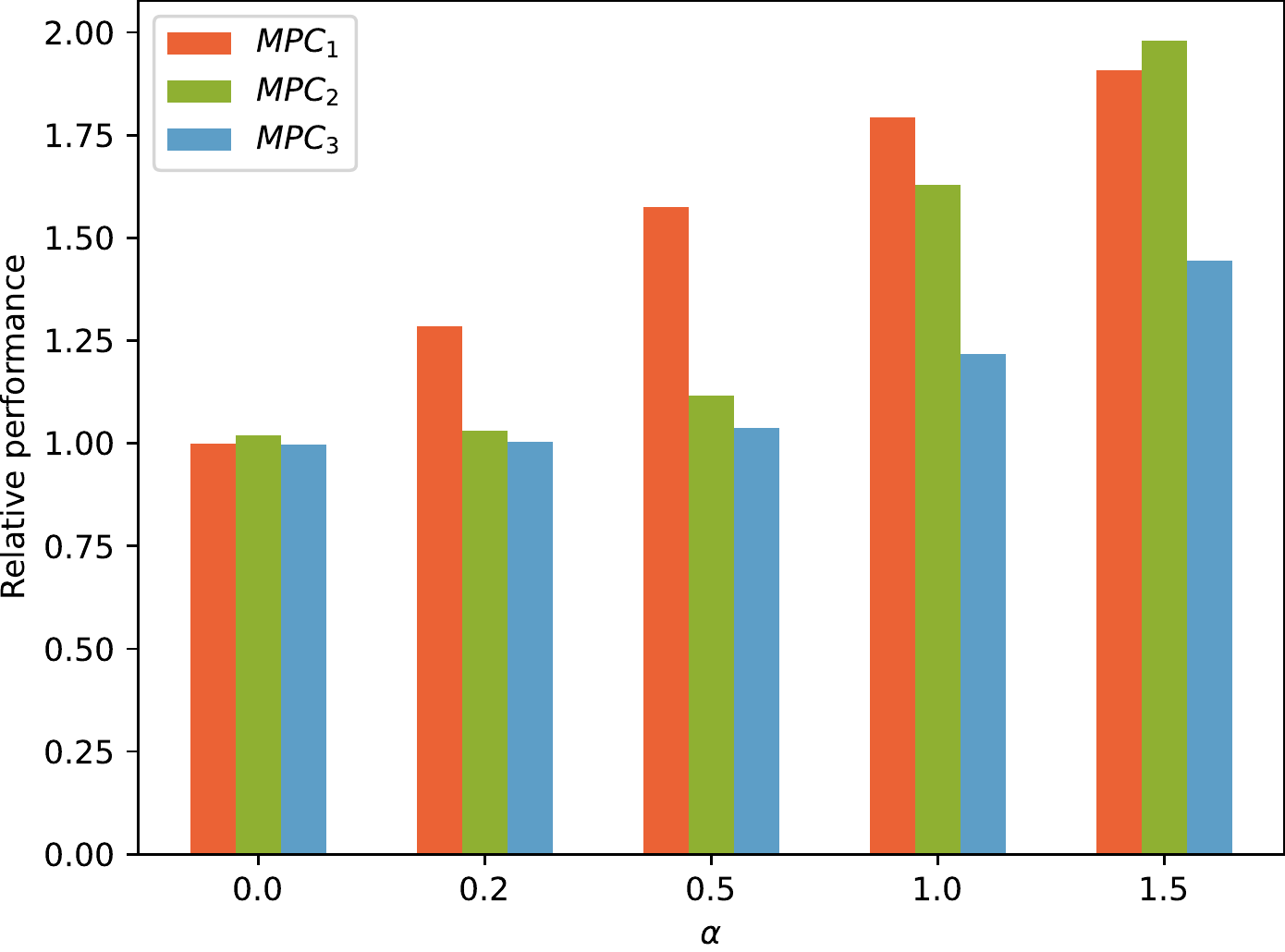}
\caption{Average performance of the MPC schemes relative to MPC with the ground truth model ($\mathit{MPC}_1$ with $\alpha = 0$) over $10$ learning trails with different seeds for pendulum swing-up task. }
\label{fig:perf_pendulum}
\end{figure}

Figure \ref{fig:perf_pendulum} shows the average of relative performance for different MPC formulations. Similar to the first example, $\mathit{MPC}_1$ starts to degrade quickly with high uncertainties in the model dynamics. Whereas, the learned MPC schemes in $\mathit{MPC}_2$ and $\mathit{MPC}_3$ manage to perform better and are able to compensate for the plant-model mismatch with $\alpha \leq 1$. Even for complex value functions learned with machine learning tools, our approach can learn the stage cost parameters in $\mathit{MPC}_2$ to compensate for inaccuracies in the given model. However, for larger model errors such as $\alpha \geq 1$, the quadratic cost in \eqref{eq:st5} cannot accurately capture $L_{\vect \theta} (\vert s, a)$ in \eqref{eq:st3}. Whereas additionally learning the model parameters in $\mathit{MPC}_3$ helps to further improve the MPC performance. However, we observe that since the value function approximation is built using finite data, $\mathit{MPC}_3$ still can not find a high-performing MPC scheme when initialized with highly inaccurate model dynamics. 

\subsection{Cartpole Swing-Up and Balancing}
The cartpole system consists of a wheeled cart of mass ($m_c=0.2$) which can freely move on a rail with a friction coefficient of ($\mu_f=0.5$), and a pole of mass ($m_p=0.2$) and length ($l=0.5$) is hinged to cart on a friction-less joint. The pole can swing freely around the hinged joint. The control input, $u$ to the system is the force exerted on the cart along the rail with $u\in[-2,2]$. The goal of the control task is to swing the pole to an upright position as quickly as possible and balance it in the upright position.

The dynamics equations for a cartpole system are given as
\begin{equation}\label{eq:cartple_system}
    \begin{gathered}
    \ddot{\phi}=\frac{g \sin \phi+\cos \phi\left(\frac{\mu_f \dot{x} -u-m_p l \dot{\phi}^2 \sin \phi}{m_c+m_p}\right)}{l\left(\frac{4}{3}-\frac{m_p \cos ^2 \phi}{m_c+m_p}\right)} \\
    \ddot{x}=\frac{u - \mu_f \dot{x} + m_p l\left(\dot{\phi}^2 \sin \phi-\ddot{\phi} \cos \phi\right)}{m_c+m_p}.
    \end{gathered}
\end{equation}
where $\phi$ is the angular deviation of the pole w.r.t the vertical position, and $x$ is the horizontal displacement of the cart. The system state-space $\vect s$ is given as $\vect s = [x, \dot{x}, \phi, \dot{\phi}]$ with $\phi$ normalized to be in $(-\pi, \pi]$. The corresponding cost function is
\begin{equation*}
     L(\vect s,u) = 2x^2 + \phi^2 + 0.1\dot{x}^2 + 0.1\dot{\phi}^2 + 0.1u^2 \, .
\end{equation*}
The task is similarly discretized with a sampling time of $0.05$, the discount factor $\gamma$ is set to $0.9$, and the task length is $100$.

The MPC scheme is formulated over the discretized task with observation $\vect y = [x,\dot{x},\cos\phi,\sin\phi,\dot{\phi}]$ with the tracking goal $\vect y_{ref} = [0,0,1,0,0]$ and MPC horizon of $N=50$. The stage cost of MPC scheme over observation $\vect y$ is 
\begin{align*}
    L_{\mathit{MPC}(\vect y, u)} =& 3x^2 + 3(\cos\phi-1)^2 + \sin\phi^2 \nonumber\\
    & + 0.01\dot{x}^2 + 0.01\dot{\phi}^2 + 0.001u^2 \,.
\end{align*}
Similar to the pendulum swing-up task, the uncertain estimate of the system parameters used in the nominal MPC model are as follows,
\begin{align}
    \hat{m}_c = m_c + \alpha\ \mathcal{U}[-0.1,0.1],\quad \hat{\mu}_f = \mu_f + \alpha\ \mathcal{U}[-0.25,0.25] \nonumber\\
    \hat{m}_p = m_p + \alpha\ \mathcal{U}[-0.1,0.1],\quad \hat{l} = l + \alpha\ \mathcal{U}[-0.25,0.25] \nonumber \,.
\end{align}

\begin{figure}[h]
\centering
\includegraphics[width=7cm]{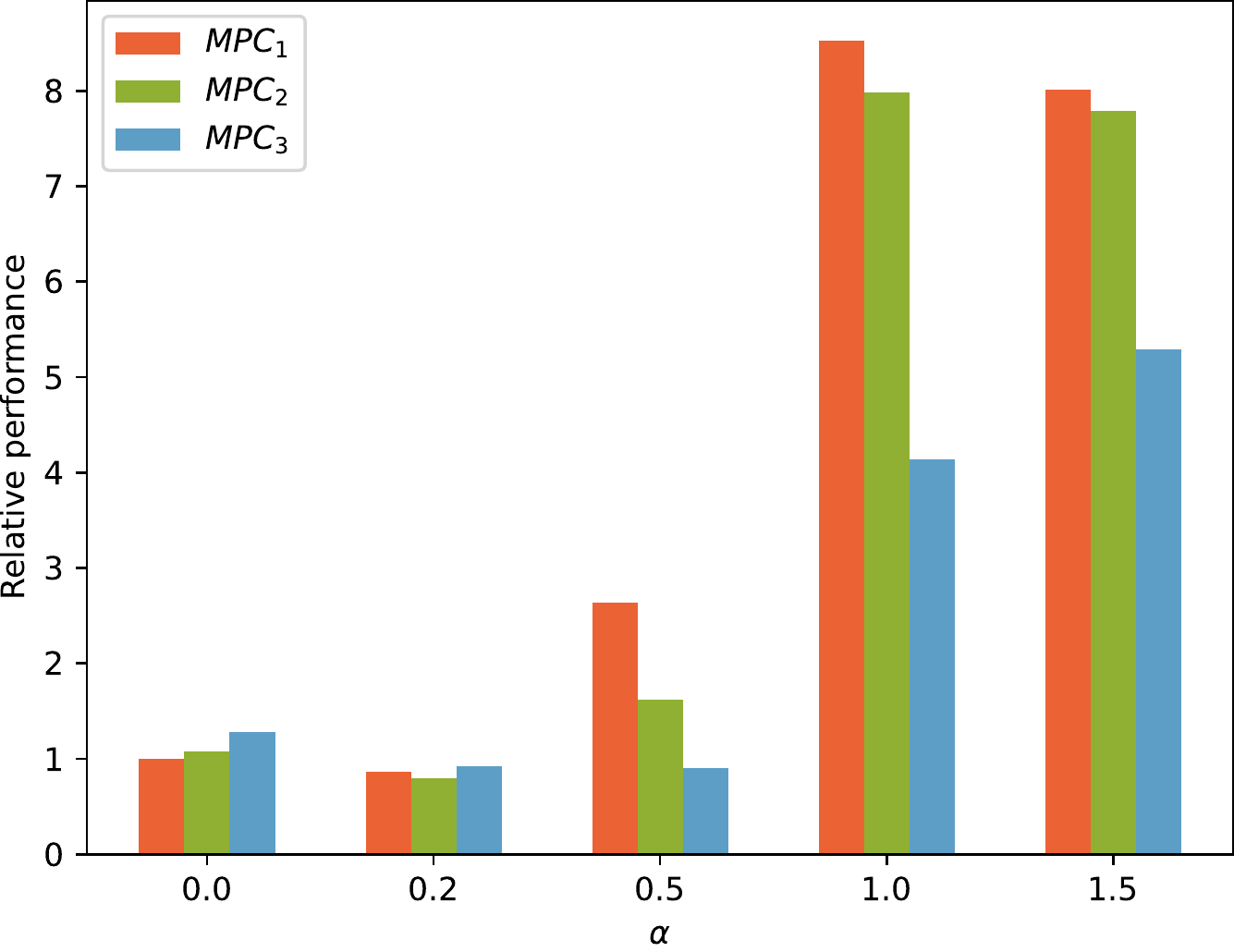}
\caption{Average performance of the MPC schemes relative to MPC with the ground truth model ($\mathit{MPC}_1$ with $\alpha = 0$) over $10$ learning trails with different seeds for cartpole swing-up and balancing task.}
\label{fig:perf_cartpole}
\end{figure}

Figure \ref{fig:perf_cartpole} shows the average relative performance of different MPC schemes for different inaccurate model dynamics. Learned MPC formulations in $\mathit{MPC}_2$ and $\mathit{MPC}_3$ similarly compensate for the model inaccuracies for $alpha < 1$ and finds a high performing MPC parameterizations. The learned MPC schemes' performance still degrades when initialized with highly inaccurate model dynamics. However, $\mathit{MPC}_3$ still out performs $\mathit{MPC}_1$, essentially, our approach still finds MPC parameters to improve performance over the baseline.

\section{Discussion} \label{sec:discussion}
The experimental results demonstrated the strength of our approach to learning an MPC scheme from data. The proposed approach outsources the task of building a good value function approximation to DNN-based RL tools, thereby bypassing the computationally expensive MPC solutions over the dataset. We essentially reduce the task of learning an MPC scheme to a simple supervised learning problem. The essence of our approach can be summarized as deriving a high-performance MPC scheme directly from data without having to endure the complexities of current RL-based MPC methods. 

Our approach learns a high-performing MPC formulation from the collected data in all three experiments, additionally compensating for inaccurate initial model dynamics effectively, irrespective of the value function complexity. Additionally from the results, we infer that the value function approximation built from a finite dataset is sufficient to extract a close-to-optimal MPC scheme using \eqref{eq:update_rule}. We observe that, by only learning the stage cost parameters, the resulting MPC scheme could correct for inaccurate model dynamics. However, cost learning can not adequately capture $L_{\vect \theta} (\vect s, \vect a)$ in \eqref{eq:st3} for highly inaccurate model dynamics.
Whereas, learning both stage cost and model parameters in an MPC scheme provides high flexibility to minimise the supervised loss in \eqref{eq:update_rule}, thereby obtaining a learned MPC scheme with close-to-optimal performance even while initialised with highly inaccurate models. 

An important point to note here is that we directly coupled the model parameter to close-loop MPC performance through \eqref{eq:update_rule}. This is a fundamentally different approach to learning the model parameters as compared to the classical approach of model learning where the parameters are adjusted for prediction accuracy.
In the context of learning model parameters, it has to be noted that the central result in \eqref{eq:update_rule} holds when optimal policy $\vect \pi_* (\vect s)$ can stabilize the model dynamics $f_{\vect \theta}(\vect s, \vect a)$. However, further investigations need to be carried out to understand the performance of our approach when this condition fails. 

In the experiments, we approximated $L_{\vect \theta}(\vect s, \vect a)$ in \eqref{eq:st3} with a quadratic cost parameterization. But such simple cost functions fail to fully capture $L_{\vect \theta}(\vect s, \vect a)$ for complex value functions and highly inaccurate model dynamics. This is evident from the results presented in Section \ref{sec:exp}. Limitations on cost function structure are a typical bottleneck in MPC and pose a similar challenge to our approach. However, it can be addressed by using rich function approximations as the stage cost in MPC schemes, such as convex neural networks in \citep{katrine2022convex}. 

Additionally, we fixed the terminal cost as $2$ norm of tracking error in the simulated examples and had a long MPC horizon to downplay its effect on MPC solutions. According to theorem \ref{th:rlmpc}, the optimal parameters $\vect \theta^*$ are such that terminal cost is the optimal value function $V^*(\vect s)$. However, a simplified value function approximation can be learned from $V_{\vect \phi}(\vect \phi)$ and further considered as a terminal cost for the learned MPC scheme. Finally, even though a rich data set is required for building a good value function approximation, a local value function estimate can be built from a sparse data distribution. Such a local value function could be used to nudge the MPC parameters toward better performance iteratively, which needs further scrutiny along the lines of the policy iteration class of methods.

\section{Conclusion}\label{sec:con}
We present an approach to formulate performance-oriented MPC schemes directly from data using RL methods. We essentially reduced the problem of deriving a high-performance MPC scheme to a supervised learning problem using a value function approximated from data. The proposed approach derives an MPC scheme in an offline way from big data, bypassing the complexities of solving MPC schemes or having to interact with the real system. Evaluations of our approach to the simulated experiments show promising results by learning a near-optimal MPC scheme from the collected dataset. In further work, we aim to apply our approach to datasets from real-world applications.


\bibliography{root}

\end{document}